\documentclass[prl,wordcount,amsfonts,amssymb,amsmath,twocolumn,nofootinbib]{revtex4}
\usepackage{epsfig}
\date{\today}
\begin{document}

\title{Zooming in on discrete space}
\author{Daniel A.\ Turolla Vanzella}\email{vanzella@ifsc.usp.br}
\affiliation{Instituto de F\'\i sica de S\~ao Carlos,
Universidade de S\~ao Paulo, Caixa Postal 369, CEP 13560-970, 
S\~ao Carlos, SP,  Brazil}
\affiliation{Institute for Quantum Optics and Quantum Information (IQOQI), 
Austrian Academy of Sciences, Boltzmanngasse 3, A-1090 Vienna, Austria\footnote{While on a sabbatical leave.}}

\begin{abstract}
Although we lack complete understanding of
quantum aspects of gravitation, it is usually
agreed, using general arguments, that 
a final quantum gravity theory will endow
space and time with some (fundamental or 
effective) notion of
discreteness. This granular character is 
supposed to lie on space and time
scales of $l_P \sim 10^{-33}$~cm and $\tau_P\sim
10^{-42}$~s, respectively---the Planck scale---, 
far beyond any hope of
direct assessment. Here, by modeling  displacements 
of particles
on a discrete underlying space
as Poisson processes, we speculate on the 
possibility
of amplifying the effects of
space discreteness (if existent) by {\it several}
orders of magnitude, 
using the statistical variance 
of correlated displacements of
particles/systems with very different masses. Although
still out of reach by current technology,
the analysis presented
here suggests that it may be possible
to 
see hints of space(time) discreteness 
at larger scales than one would usually expect.

\end{abstract}

\maketitle

The comprehension that the most basic concepts 
of space and time are fundamentally 
dynamical---ultimately giving rise to the phenomenon of gravity, as ruled, e.g., by General Relativity---is certainly one of the most striking paradigm shifts in Physics. Possibly, it is also at the heart of the difficulty in formulating a gravity theory which fully encompasses the quantum principles which seem to rule all other 
fundamental
aspects of Nature. Although a complete quantum gravity theory continues elusive after a search of more than a century, there are some features  which are generally 
believed that will be true
in such a theory. One of them is the (fundamental or effective) discreteness of space and time at the Planck 
scale---$l_P \sim 10^{-33}$~cm and 
$\tau_P \sim 10^{-43}$~s, respectively. Unfortunately, 
directly assessing such diminutive scales through
standard
Particle Physics experiments---which would 
involve particle
processes with energies of order $10^{19}$~GeV (the 
Planck energy), some ${15}$ orders of magnitude beyond
LHC's energy scale---seems a hopeless task. 
A (loose)
parallel can perhaps 
be drawn with the ``atomistic theory'' at the end of the
$19^{\rm th}$ century, when  
the physical reality
of the hypothetical entities called {\it atoms}
and {\it molecules}, which were so useful for 
stoichiometric calculations in Chemistry, was beyond direct
verification due to their
(alleged) tiny dimensions. Notwithstanding, 
the {\it statistical} properties of (the 
existence of) a large number of these invisible 
entities were directly linked, by Einstein, to the {\it observed} phenomenon of the Brownian 
motion~\cite{BM}---hence, giving observational support to the
reality of those invisible entities.
Could statistics come
to our help, again, regarding the existence
of a discrete spatial scale? Here, we try to argue that this may indeed be
possible. (We use natural units, in which $c = \hbar = 1$.)

In order to illustrate the general idea, consider a one-dimensional space discretized in cells, each with size $l_P$. An ``object'' which possesses only one spatial degree of freedom on this space cannot be localized to a precision better than 
$l_P$---regardless how many cells its extension 
occupies\footnote{By definition, having only one spatial 
degree of freedom means that the entire 
configuration of the ``object'' can be characterized by a single
parameter, even though this configuration may occupy several spatial
cells.}. 
Therefore, if this object is ``forced''  to undergo a certain 
spatial displacement $d$
(e.g., 
by some ``physical law''---we
ask the reader to bear with 
us for now), it 
does not seem too far-fetched
that
this process 
might be treated
as a {\it Poisson process} for the
discrete  variable $d_N := N l_P$ ($N\in{\mathbb N}$),
with
average $\langle d_N\rangle = d$. 
This constitutes the first basic assumption underlying our investigation.\footnote{The analysis and conclusions presented here
could be adapted to processes following statistical distributions other than Poissonian.}

The assumption above means that if the process of
``forcing'' the same object to undergo  the same spatial translation
$d$ is repeated a large number of times, there should exist 
an {\it inevitable statistical variance} in the determination of 
$d_N$, associated to a standard deviation $\Delta d_N = 
d/\sqrt{N} = \sqrt{l_P d}$. 
This is certainly a natural consequence of describing 
position as a
discrete  variable on an underlying discrete space. And the
standard-deviation
expression can be used to illustrate well
the difficulty in {\it observing} this statistical effect of space 
discreteness if $l_P$ is
exceedingly
small. For in order to detect such an effect, one should be able 
to {\it measure} the variable $d$ to an experimental
precision $\Delta d$ better than
$\Delta d_N$  (i.e., $\Delta d <\Delta d_N$).
For instance:
given an experimental  {\it accuracy} $a := \Delta d/d$ (i.e., 
relative precision) with which $d$ can be measured in an experiment,
$\Delta d <\Delta d_N$ implies $N<1/a^2$, which, in  turn, 
leads to $d<l_P/a^2$ and $\Delta d <l_P/a$.
Therefore, even with extremely accurate displacement
measurements, 
say $a \sim 10^{-10}$, one would have to measure displacements
of order $10^{-15}$~m (the size of a proton) 
to a precision better than $10^{-25}$~m
in order to be sensitive to this statistical variance---recalling 
that
$l_P\approx 10^{-33}$~cm.
If instead of fixing the accuracy we  give the  precision $\Delta d$
to which an experiment  can determine $d$, say 
$\Delta d \sim 10^{-12}$~m (the Compton wavelength of an 
electron), then the condition $\Delta d <
\Delta d_N$ implies that one should be able to measure, to that
precision, displacements $d > \Delta d^2/l_P \sim 10^{11}$~m
(the radius of Venus's orbit). These figures
illustrate very clearly why statistical effects of
space discreteness would have passed unnoticed even if our basic
assumption
(of modeling displacements as Poisson processes) is 
true.\footnote{Note that the requirement is to measure a
{\it displacement} $d$ to a precision $\Delta d$. This is {\it not}
the same  as measuring a {\it distance} $d$ between two objects
to that 
same precision---the latter being {limited} 
only by the two-object's
uncorrelated, individual uncertainties,
independent of the distance between them.}

In fact, things can be even worse than discussed 
above---i.e., the statistical effect we are interested in can be even more elusive.
This is  because we  only considered ``objects'' with one spatial
degree of freedom on a given direction.
Adding spatial degrees of freedom ($f$) 
 to the
analysis above can  have 
two effects. On the positive---and practical---side, 
it allows us to
consider ``objects'' with arbitrary masses $m$, 
beyond the ones of the 
known
elementary particles. And even for elementary particles, the assumption 
of one or few spatial 
degrees of freedom ($f\sim 1$)
is not in general
realistic, due to the supposedly continuous nature of their quantum-mechanical
wave functions. 
Nonetheless, $f\sim 1$ is possible to be attained
for
 specific quantum states---e.g., 
ground states in confining potentials. 
So, as an idealization, we shall consider, here, 
that
it is {\it possible} to have $f\sim 1$ for ``objects'' up to a
certain
mass/energy scale $\mu$---which we simply call ``particles''---above which $f$ may 
increase with 
$m$ (e.g., with a power-law dependence such as $f  =
(m/\mu)^{\alpha} $,  $\alpha> 0$ and $m>\mu$).

On the other hand, the negative effect (for observational purposes) 
of
adding spatial degrees of freedom is that it makes 
the measurement conditions more stringent.
This is because, typically, $f >1$ improves the precision to which, 
{\it in principle}, the object can be 
localized ``on average'' on a discrete grid---for
instance, its center of mass, which is the spatial degree
of freedom which is associated
to the whole mass $m$. 
This makes  the 
{\it effective} discreteness scale for the whole mass $m$, 
 $l_{\text{\it eff}}$,
even smaller than the Planck length if $f\gg 1$, which, in turn,
makes the ``effective number of cells'' 
$N_{\text{\it eff}} = d/l_{\text{\it eff}}$, for a given displacement $d$,
even larger---thus suppressing
statistical fluctuations.
For concreteness sake, we model
$l_\text{\it eff} = l_P/f^{\beta}$, with $\beta > 0$.
We leave our expressions in terms of 
$l_\text{\it eff}$, which is {\it the}
quantity on which our discussion depends. But 
we ask the reader to bear in mind this conjectured dependence of $l_\text{\it eff}$
on $f$ and of this latter on $m$---which
leads to $l_\text{\it eff} = (\mu/m)^{\alpha\beta} l_P$ for $m>\mu$ (while  $l_\text{\it eff} = l_P$ for $m\leq \mu$). 
When illustrating numerical values,
we consider the case $f\propto m$ (i.e., $\alpha = 1$) and 
$l_\text{\it eff}\propto 1/\sqrt{f}$ (i.e., $\beta = 1/2$)
for $m>\mu$,
which seem reasonable as a toy model (the latter due to the
assumption of
statistical independence of the spatial variances associated to 
each
degree of freedom).

Thus, repeating the same analysis performed earlier for one degree
of freedom but now for $f>1$ degrees of freedom---with the additional
constraint that, in order to be unambiguously 
{\it measurable}, in principle, the 
statistical deviation 
$\Delta d_N$ should be considerably larger than 
$l_P$\footnote{The reader may well question why we impose
$\Delta d_N \gg l_P$ instead of $\Delta d_N\gg l_\text{\it eff}$. We may 
justify this by pointing out that the fact that $l_\text{\it eff}$ is the effective
discrete spatial scale for purposes of inferring statistical properties 
of displacements of the object, this does not  necessarily mean  that our 
``displacement measuring device'' is not bound by the fundamental
scale $l_P$. And in any case, the condition 
$\Delta d_N \gg l_P$ is stronger than $\Delta d_N\gg l_\text{\it eff}$,
in the sense that by imposing the former we also guarantee the latter.} and $\lambda_C=1/m$ (the Compton wavelength of the object with 
mass $m$)---, 
the condition that space discreteness should lead 
to an (in principle) observable
statistical variance
in displacement measurements (i.e., $l_P,\lambda_C \ll \Delta d < \Delta d_N 
= \sqrt{l_{\text{\it eff}}\,d}$) imposes
\begin{eqnarray}
d\gg
\left\{
\begin{array}{ll}
\lambda_C^2/l_\text{\it eff} &, m\leq m_P\\
l_P^2/l_\text{\it eff} &, m>m_P
\end{array}
\right.
\label{dmin}
\end{eqnarray}
and, consequently,
\begin{eqnarray}
\frac{\Delta d}{d}\ll
\left\{
\begin{array}{ll}
l_\text{\it eff}/\lambda_C &, m\leq m_P\\
l_\text{\it eff}/l_P &, m> m_P
\end{array}
\right.,
\label{Deltadmin}
\end{eqnarray}
where $m_P :=1/l_P\approx 10^{19}$~GeV is the Planck mass.

\begin{figure*}
\includegraphics[width=\textwidth]{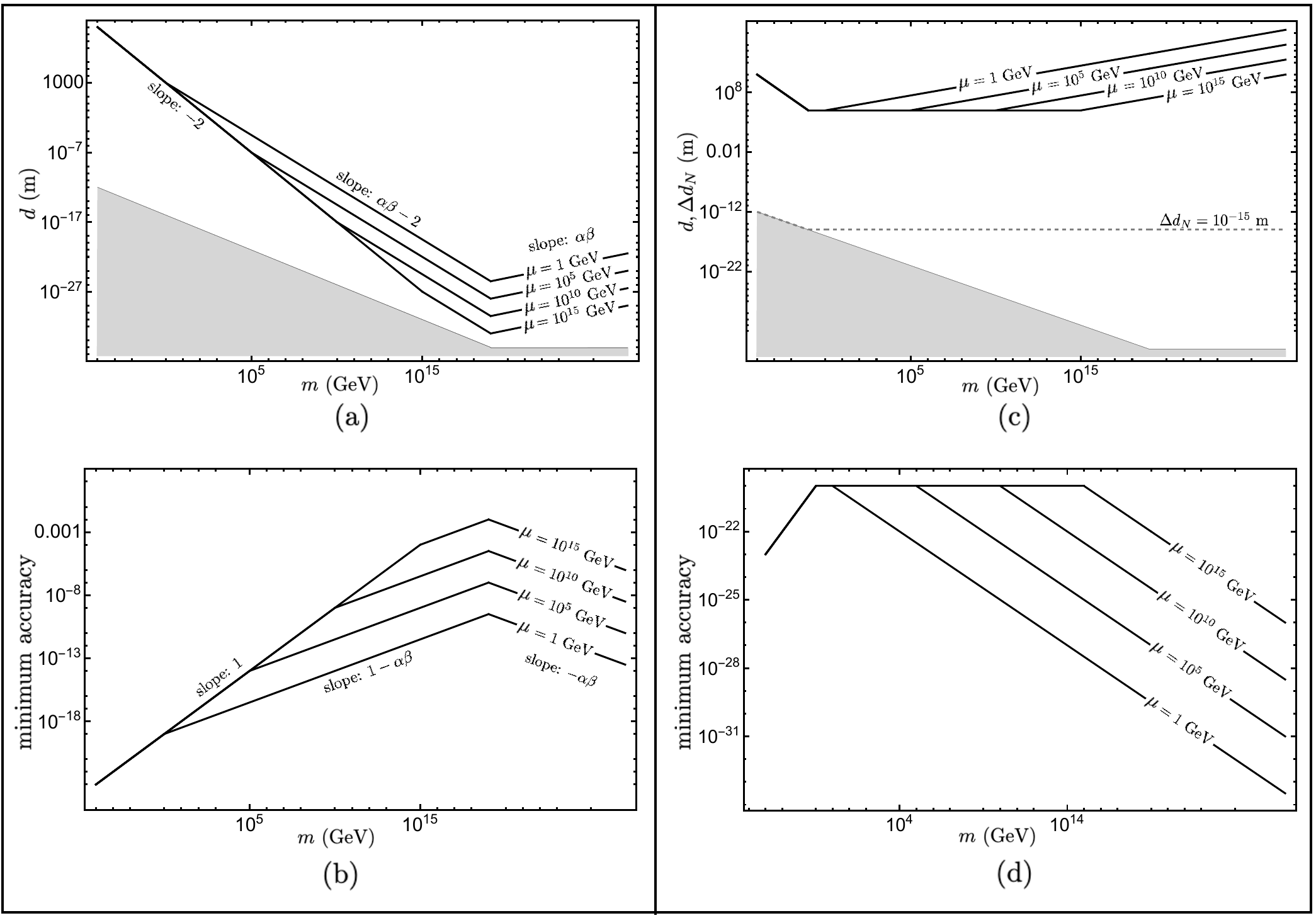}
\caption{(a)~The curves represent the r.h.s.~of Ineq.~(\ref{dmin}) 
assuming $l_\text{\it eff} = l_P(\mu/m)^{\alpha\beta}$ (with $\alpha =1 $ and $\beta = 1/2$), for different values of
$\mu$. These are the lower-bound values of displacement $d$ for which
statistical variance due to space discreteness
is {\it in principle} measurable (i.e., 
$\Delta d_N \gg \lambda_C, l_P$). The gray region represents distances 
smaller than $\lambda_C$
or $l_P$. (b)~The curves represent the r.h.s.~of Ineq.~(\ref{Deltadmin}),
again assuming $l_\text{\it eff} = l_P(\mu/m)^{\alpha\beta}$ (with $\alpha =1 $ and $\beta = 1/2$), for different values of
$\mu$. These are the worst-accuracy bounds needed to observe statistical variance
in measurements of spatial displacements of a mass $m$ due to space discreteness.
(Recall that better accuracy means lower values in the vertical axis.)
(c)~The curves represent the
lower-bound values of displacement $d$ for which
statistical variance due to space discreteness
can in principle
be measured to a precision $\Delta d \approx 10^{-15}$~m (assuming $l_\text{\it eff} = l_P(\mu/m)^{\alpha\beta}$, with $\alpha =1 $ and $\beta = 1/2$).
The gray region represents distances 
smaller than $\lambda_C$
or $l_P$.
(d)~The corresponding worst-accuracy bounds
needed in the displacement measurements
of Fig.~\ref{fig:1}(c).
}
\label{fig:1}
\end{figure*}

Just for the sake of illustrating how $f>1$ makes the task of 
detecting statistical variance coming from discrete space more 
difficult, we
plot, in 
Fig.~\ref{fig:1}(a), 
the right-hand side (r.h.s.)~of 
Ineq.~(\ref{dmin}) (for different values of $\mu<m_P$) assuming the
dependence $l_\text{\it eff} = (\mu/m)^{\alpha\beta}l_P$ conjectured
above, with $\alpha = 1$ and $\beta = 1/2$. 
The curves
represent (for different values of $\mu$) the (underestimated) 
lower bounds 
on the displacement $d$ of an object with mass $m$ for which
statistical variance due to space discreteness would {\it in principle} be
measurable. 
In Fig.~\ref{fig:1}(b), 
we plot the corresponding 
worst-accuracy bounds [the r.h.s.~of Ineq.~(\ref{Deltadmin})] 
for measurements of $d$ to
be sensitive
to statistical variance. (Recall that ``better
accuracy'' means smaller values on the
vertical axis.)
We  see that, regardless the (fixed) value of $\mu$
(and for $0<\alpha\beta<1$),
objects with mass $m\sim m_P\approx 10^{19}$~GeV 
are the ones which, 
in principle, would require {\it less accuracy}
in their displacement measurements to still 
be sensitive to
variance due to space discreteness. But this is simply because
these are the objects whose best-possible localization scale
($\sim\lambda_C\sim l_P$) 
is the
closest to the (effective) 
discreteness scale ($\sim l_\text{\it eff}$).
Obviously,
this is completely useless for any practical purposes, 
since the {\it absolute
precision} ($\Delta d$) needed  to be indulged by this 
``low accuracy''
is close the Planck scale itself. 

In Fig.~\ref{fig:1}(c), instead of privileging accuracy, 
we plot the minimum displacement necessary if the {\it absolute
precision}
is limited by, e.g., $\Delta d \approx 10^{-15}$~m. We see that, in 
this case,
observing statistical variance due to space discreteness would
involve measuring, to a precision comparable to the size of 
an atomic nucleus, ``identical'' 
displacements 
varying from a few hundred kilometers (for masses
$m\sim \mu$) 
up to the distance to the Oort Cloud ($\sim 10^{15}$~m, 
for $\mu\sim 1$~GeV and $m\sim m_P$), beyond the limits of our 
Solar System.
This ludicrous situation 
is well reflected in 
the  worst-accuracy bounds needed in these cases,
plotted in Fig.~\ref{fig:1}(d).

\medskip
So far, all we have done, using quantitative 
arguments based on our simple model, is corroborating
 the common wisdom that
effects of an hypothetical  granular nature of space at scales $l_P\sim 10^{-33}$~cm
are
well beyond any hope of direct observation.
With this resigned view, we ask the reader to consider the following idealized
situation.

\begin{figure}
\includegraphics[width=0.45\textwidth]{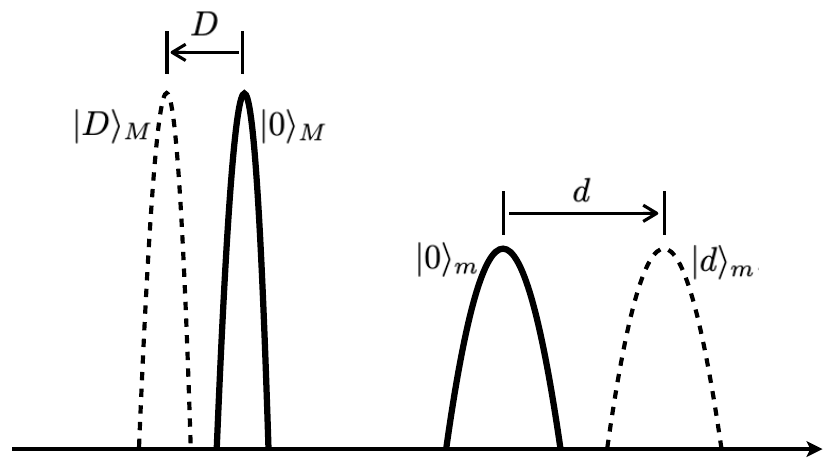}
\caption{Pictorial representation of displacements $d$ and $D$ due to interaction
between the particle with mass $m$ and the ``object'' with mass $M$. Conservation 
of the mass dipole moment of an isolated system leads to $md+MD = 0$.
}
\label{fig:2}
\end{figure}

A particle with mass $m$ and an ``object'' (i.e., possibly having $f\gg 1$)
with mass $M\gg m$ are initially at rest w.r.t.~each other, constituting
an {\it isolated} system.
Let $|i\rangle = |0\rangle_m|0\rangle_M$ be the initial state of the system, where $|0\rangle_m$ and $|0\rangle_M$ represent the initial
(uncorrelated) peaked position states of the particle and the object, respectively; let $\sigma_m$ and $\sigma_M$ be the corresponding  spatial
uncertainties (in principle, limited by $\sigma_m >1/m,l_P$ and $\sigma_M 
>1/M,l_\text{\it eff}$).
Now, suppose 
interaction takes place between them,
in such a way that the state of the particle is displaced 
by a certain distance $d$---characterized 
by the final state
$|d\rangle_m$.
Conservation of total
momentum of an isolated system imposes that its center of mass
(or, equivalently,
its mass dipole moment)
cannot change in the initial, inertial rest frame. Therefore, the
final position state $|d\rangle_m$ 
of the particle
{\it must} be correlated to a final 
position state $|D\rangle_M$ of the object, with $md+MD = 0$,
so that the final state of the system 
is described by 
$
|f\rangle = |d\rangle_m|D\rangle_M
$ (see Fig.~\ref{fig:2}).
In principle, there is no reason for the uncertainty in the position of 
the
center of mass  
of the system to increase appreciably during 
this
process; i.e., the {\it inherent} uncertainties of the
final states $|d\rangle_m$ and $|D\rangle_M$ in each 
realization of the experiment can, in principle, be the same as
the ones of the initial states $|0\rangle_m$ and 
$|0\rangle_M$,
respectively. 

\begin{figure}
\includegraphics[width=0.45\textwidth]{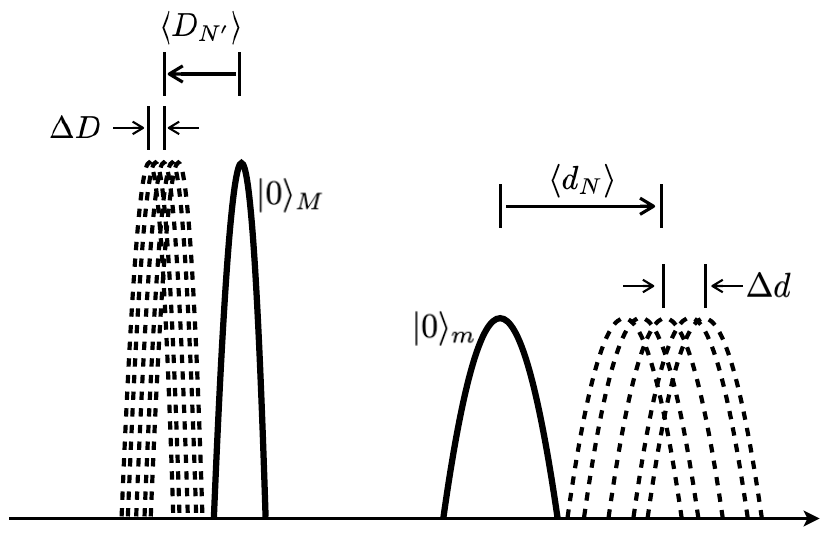}
\caption{Pictorial representation showing several processes ``identical'' to the one represented in 
Fig.~\ref{fig:2}. The
statistical variance of the discrete variable $D_{N'}$ is transferred, 
amplified,
to the variable $d_N$ due to the constraint $md_N + M D_{N'} = 0$.
}
\label{fig:3}
\end{figure}

Consider, now, repeating this same 
{\it identical} process a 
large number of times. {\it If} $d$ (and $D$)
were a continuous variable,
the statistical variance in 
the position/displacement measurements of the particle should
reflect {\it only} the 
uncertainty $\sigma_m$ 
associated to $|d\rangle_m$.
However, if we take into consideration that the spatial  displacements
should be described by the discrete variables $d_N$ and $D_{N'}$,
with $md_N+MD_{N'} = 0$ for {\it each} run (see Fig.~\ref{fig:3}), we end up with 
two {\it coupled} Poisson
processes for which the standard deviations
$\Delta d$ and $\Delta D$ 
are tied together by
\begin{eqnarray}
m \Delta d = M
\Delta D \;\Leftrightarrow\;
\frac{\Delta d}{|\langle d_N\rangle|} = \frac{\Delta D}
{|\langle D_{N'}\rangle|}.
    \label{DdNDDN}
\end{eqnarray}
As a consequence, the {\it accuracy} to which the displacements $d$ and $D$
can be determined is {\it limited} by the one which is most affected by
statistical variance. From Fig.~\ref{fig:1}(b), we see that
statistical variance
should be most important, in principle, 
for masses around the Planck mass $m_P$---as previously explained,
due to the
fact that these are the masses whose best-possible localization scale is 
the closest to
the discreteness scale $l_\text{\it eff}$.\footnote{If we relax the condition
$\Delta D\gg l_P$ used to obtain Fig.~\ref{fig:1}(b) and consider the weaker one, $\Delta D \gg 
l_\text{\it eff}$---which seems justifiable, here, since we are not assessing
{\it directly} the displacement $D$ (see previous footnote)---then the 
higher the mass $M$, the worse
the accuracy $\Delta D/D$---{\it provided} 
$f$ does {\it not} grow linearly or faster with 
$M$---and the easier it would be to observe effects of statistical variance
on $D$. However, the hypothesis that $f$ does not grow linearly or faster
with $M$ (i.e., $\alpha<1$) does impose a nontrivial condition which is likely to
be false or experimentally untenable for large enough masses. Therefore, we stick to 
the stronger condition $\Delta D \gg l_P$.} So, by choosing
$M\approx m_P$, the statistical variance
in the determination of $D$ ``contaminates'' the determination of 
(or gets transferred to) $d$ via
Eq.~(\ref{DdNDDN}), which
leads to
\begin{eqnarray}
\Delta d &=&\frac{\Delta D}{|\langle D_{N'}\rangle|} |d| = 
\frac{|d|}{\sqrt{N'}}
= \sqrt{\frac{l_\text{\it eff}(M)}{|D|}}\, |d|
\nonumber \\
&=&
\sqrt{\frac{M l_\text{\it eff}(M) |d|}{m}},
    \label{Deltad}
\end{eqnarray}
where, to be clear, $l_\text{\it eff}(M)$ stands for the effective
scale of space discreteness for the object with mass $M$ (taking into
account its $f(M)$ spatial degrees of freedom).
If we want this ``contamination'' 
arising from space discreteness to be in principle
measurable for the particle with mass $m \ll M \approx m_P$, we
need to impose $\Delta d \gg  1/m,\sigma_m$
(in addition to $\Delta D\gg l_P$; see
previous footnote). 
Assuming, just to make comparisons simpler, that $\sigma_m$ can be made of the same order as $\lambda_C = 1/m$,
this leads to
\begin{eqnarray}
d\gg 
\left\{
\begin{array}{ll}
 (m_P/M) [l_P/l_\text{\it eff}(M)]\,\lambda_C&, \;\;M<m_P\\
(M/m_P)[l_P/l_\text{\it eff}(M)]\lambda_C  &, \;\;M\geq m_P
\end{array}
\right. .
    \label{finalcond}
\end{eqnarray}

\begin{figure*}
\includegraphics[width=0.6\textwidth]{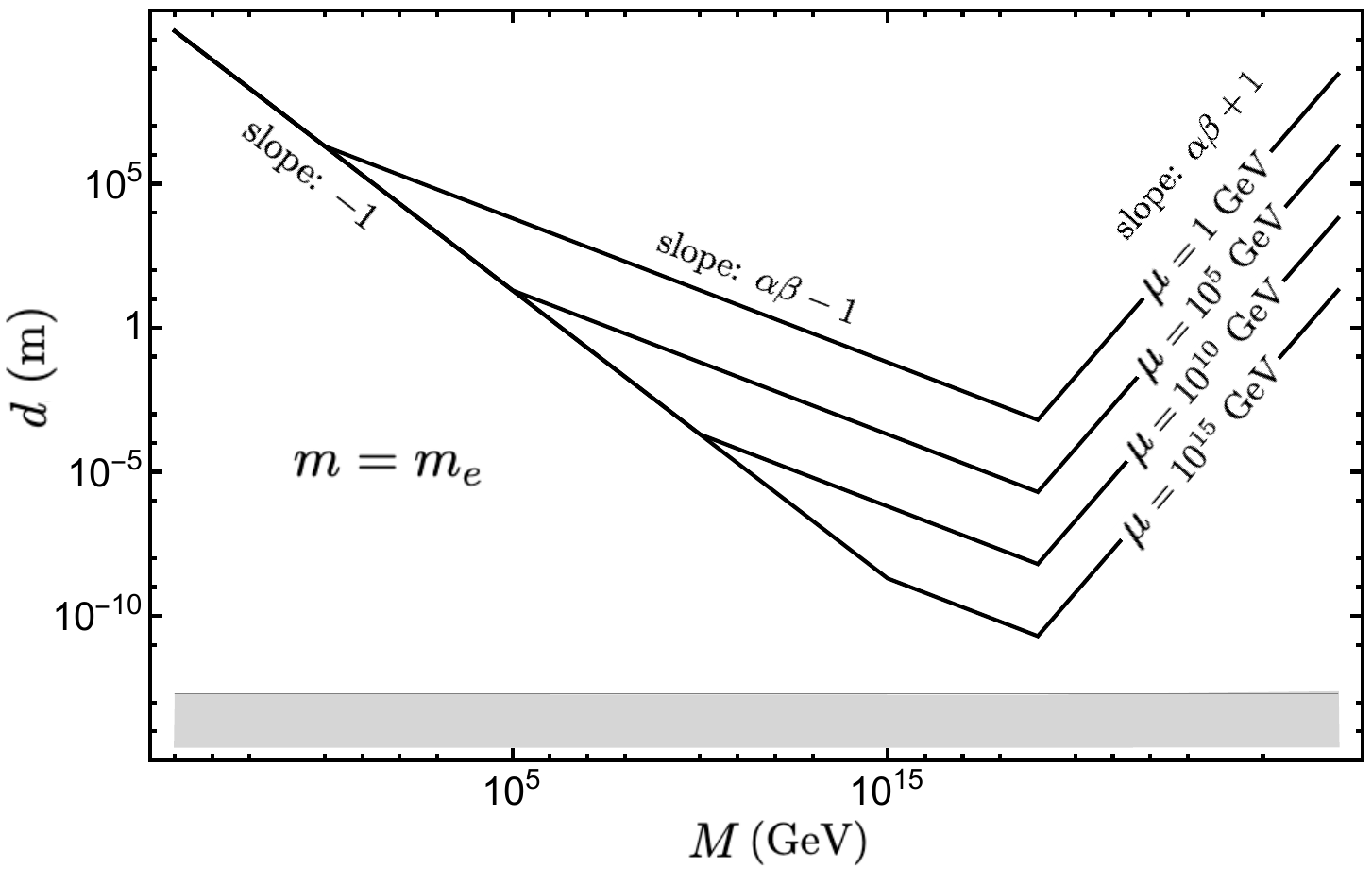}
\caption{The curves represent the r.h.s.~of Ineq.~(\ref{finalcond})
(assuming $l_\text{\it eff} = l_P(\mu/M)^{\alpha\beta}$, with $\alpha =1 $ and $\beta = 1/2$). They
correspond to the lower-bound values of displacement of an
electron-mass particle in order to statistical
variance, due to an underlying discrete
space, to be {\it in principle} observable when the
displacement is due to interaction with an ``object'' with mass $M$. (The gray area marks the region where
$d < \lambda_C = 1/m_e$.) 
}
\label{fig:4}
\end{figure*}

The result above is to be compared with 
Ineq.~(\ref{dmin}).
The conclusion is: while independent displacements $d$ of a particle with mass $m <m_P$
must satisfy $d\gg \lambda_C^2/l_\text{\it eff}(m)$ for
 statistical variance due
to space discreteness to be {\it in principle} observable ($\Delta d_N \gg \lambda_C$),
they only need to satisfy the {\it weaker} condition 
$d\gg \left[l_P/l_\text{\it eff}(M)\right]\lambda_C$ if they are coupled, through mass-dipole-moment
conservation, to the displacements $D$ of an object with mass $M\approx m_P$.
For the sake of illustration, in Fig.~\ref{fig:4} we plot the
r.h.s.~of Ineq.~(\ref{finalcond})---under the same assumptions used to plot 
Fig.~\ref{fig:1}, namely $l_\text{\it eff} (M)= l_P/[f(M)]^\beta$ and $
f(M) ={\rm max}\{1, (M/\mu)^\alpha\}$, with $\alpha = 1$ and $\beta = 1/2$---for
$m = m_e$ (the mass of the electron) and different values of $\mu$.
We see that instead of having to measure displacements of
order $10^{11}$~m (Venus's orbital radius) 
for an electron with a precision comparable to its Compton wavelength,
now we ``only'' need to measure displacements of about $1$~cm to that same precision
(in case $\mu = 1$~GeV)
if this displacement is due to interaction with a Planck-mass 
object---thus turning an astronomically difficult problem into a down-to-Earth
(though still difficult) task, literally.

\medskip
In conclusion, here we have explored the idea that
spatial discreteness alone should naturally lead to  statistical variance 
of the spatial displacement of an object.
The question we investigated, in this scenario, was: can this statistical variance be observed and, hence, provide a signature of the existence of a minimal length scale? We have shown that although this is a hopeless task for independent
displacements (see Fig.~\ref{fig:1}),
interaction between particles with different masses, modeled as
{\it coupled}
Poisson
processes, can {\it amplify}, by several orders of magnitude,
statistical variance across the very different mass scales.
Our analysis here should be seen as a mere ``proof of concept.''
Obviously,
several experimentally-relevant issues and subtleties 
were neglected here, as, e.g.,
the experimental
conditions under which the particle's and object's states  
would satisfy the {\it key condition}
$$\frac{M \,l_\text{\it eff}(M)}
{m \,l_\text{\it eff}(m)}\gg 1$$ 
---which
is necessary for
 Ineq.~(\ref{finalcond}) to represent a significantly  
 less stringent
 condition than Ineq.~(\ref{dmin})---not to mention
 the obvious requirement that $\Delta d$ given by Ineq.~(\ref{Deltad})
 should not be buried under $\sigma_m$ ($\Delta d \gg \sigma_m$). In fact, distinguishing $\Delta d$ coming from 
Eq.~(\ref{Deltad}) from $\sigma_m$ coming from the spread 
of the
wave function associated to $|d\rangle_m$ seems  
a 
delicate experimental
task.
Another (over)simplification considered here regards the final
state $|f\rangle = |d\rangle_m|D\rangle_M$ of the system. 
Since the idea is to amplify tiny statistical variances of 
already tiny and possibly directly-unobservable
displacements $D$,
it seems much more reasonable to consider 
final states of the {\it entangled} form
\begin{eqnarray}
|f\rangle = \int_{\mathbb R} dD f(D) |d(D)\rangle_m |D\rangle_M,
    \label{genf}
\end{eqnarray}
with $d(D) = -M D/m$ and $f(D)$ a complex function, satisfying
$\int dD |f(D)|^2 = 1$, 
which depends on the details of the experiment. Although this does not
invalidate the core idea that measurements of $d$ can exhibit an inevitable
(and measurable)
statistical variance coming from the effects of space discreteness
on the variable $D$, it certainly modifies the details of the analysis.\footnote{The object's 
displacement $D$ would play the role of a {\it nonlocal} hidden variable for the state of the
mass $m$.}
Moreover, 
considering the perspective that, in the not-so-distant future, 
very sensitive
experiments will likely be able to 
generate entanglement of position states
of mesoscopic massive particles---e.g., through Coulomb
or gravitational~\cite{Grav}
interaction---it is interesting to consider cases
where
$$
|f\rangle \approx \frac{1}{\sqrt{2}}\left(|\ell\rangle_m|R\rangle_M+
|r\rangle_m|L\rangle_M\right),
$$
with $|\ell\rangle_m$, $|r\rangle_m$ and $|L\rangle_M$, $|R\rangle_M$ being
peaked position states of the masses $m$ and $M$, respectively. Although the main
goal of these
planned experiments is observing elec\-tro\-magnetic\-ally- and gravity-induced position
entanglement between particles, it would be interesting to analyze how close 
they would be, in terms of sensitivity,
to seeing hints of an underlying discrete space,
in the sense discussed here.

\acknowledgments
\subsection{Acknowledgments}
This work was developed while on a sabbatical leave
at the Institute for Quantum Optics and Quantum Information
(IQOQI-Vienna) of the Austrian Academy of Sciences. The author 
acknowledges
partial financial support from
the S\~ao Paulo State Research Foundation
(FAPESP) under grant no.\ 2023/04827-9. 
The author thanks the IQOQI, the University 
of Vienna, and
\v Caslav Brukner and his group for the
stimulating environment and their hospitality. The author also thanks
Gerard Higgins for inspiring conversations 
about  state-of-the-art quantum experiments and Jeremy Butterfield for
valuable suggestions on the text.

\end{document}